\documentclass[
aps,pre,
reprint,showpacs,amsmath,amssymb,
groupedaddress
]{revtex4-1}
\usepackage{graphicx,times,tikz}
\pdfoutput=1

\begin{document}

\title{Subwavelength Total Acoustic Absorption with Degenerate
	Resonators}
\author{Min Yang}\altaffiliation{Contributed equally to this work}
\author{Chong Meng}\altaffiliation{Contributed equally to this work}
\author{Caixing Fu}
\author{Yong Li}
\author{Zhiyu Yang}
\author{Ping Sheng}
\affiliation{
Department of Physics, Hong Kong University of Science
and Technology, Clear Water Bay, Kowloon, Hong Kong, China
}

\begin{abstract}

We report the experimental realization of perfect sound absorption by
sub-wavelength monopole and dipole resonators that exhibit degenerate
resonant frequencies.  This is achieved through the destructive
interference of two resonators' transmission responses, while the
matching of their averaged impedances to that of air implies no
backscattering, thereby leading to total absorption.  Two examples,
both using decorated membrane resonators (DMRs) as the basic units,
are presented.  The first is a flat panel comprising a DMR and a pair
of coupled DMRs, while the second one is a ventilated short tube
containing a DMR in conjunction with a sidewall DMR backed by a
cavity.  In both examples, near perfect absorption, up to 99.7\%, has
been observed with the airborne wavelength up to 1.2 m, which is at
least an order of magnitude larger than the composite absorber.
Excellent agreement between theory and experiment is obtained.

\end{abstract}
\maketitle

Total absorption of sound using subwavelength structures or materials
has always been a challenge, since the linear dynamics of dissipative
systems dictates the fractional power to be linearly proportional to
the elastic deformation energy \cite{landau1970elasticity}, which is
negligible in the sub-wavelength scale.  To enhance the dissipation,
it is usually necessary to increase the energy density, for example,
through resonances.  However, in an open system, radiation coupling to
resonances is an alternative that can be effective in reducing
dissipation.  In previous studies, by utilizing localized
subwavelength resonances, membrane-type metamaterial
\cite{yang2008membrane,lee2010reversed,park2011amplification,park2013giant,yang2013coupled,ma2013low},
containing decorated membrane resonator (DMR) with tunable weights,
has shown efficient and flexible capability in low frequency sound
absorption \cite{mei2012dark}.  A balance between dissipation and
scattering at resonance has been found for optimum absorption
\cite{yang2015sound}.  More recently, a perfect absorber has been
realized by hybridizing DMR's two resonances through coupling via a
thin gas layer.  Through interference, waves reflected from such DMR
have been shown to completely cancel that from a reflective wall
placed a short distance (about 1/133 of airborne wavelength) behind
the DMR \cite{ma2014acoustic,yang2015sound}.  Meanwhile, the coherent
perfect absorber (CPA) in optics shows that the scattering waves at
resonance can be cancelled when another counter-propagating coherent
light wave, with specific phase and intensity, interferes with the
incident beam, thereby leading to total absorption
\cite{chong2010coherent,wan2011time,noh2012perfect,nie2014selective,li2014equivalent,piper2014total2}.
Recent efforts have also been made for its analogy in acoustics
\cite{wei2014symmetrical,song2014acoustic,cai2014ultrathin,leroy2015superabsorption}.
However, except for some theoretical attempts in acoustic
\cite{lapin2003monopole} and numerical studies in optics
\cite{piper2014total}, up to now no perfect absorber has been
experimentally realized that intrinsically eliminates all the
scattered waves, thereby realizing total absorption regardless of the
incident direction, and with no need for a control wave.

In this article, we advance the idea of creating a total acoustic
absorption unit comprising a monopole (symmetric under mirror
reflection) and a dipole (anti-symmetric) resonator that are resonant
at the same frequency.  Similar to the hybrid resonance, this
degenerate absorption unit can have two useful degrees of freedom,
inherited from the two resonances.  During a scattering event,
reflection can be eliminated by making the average impedance of the
two resonators to be impedance-matched with the background medium.
Transmission can also be eliminated if the response on the other side
is zero due to destructive interference.  The net result is a perfect
absorber that scatters no sound and that does not need a reflecting
back wall as in the hybrid membrane resonator (HMR) or another
controlling wave as in the CPA.  Owing to its subwavelength
dimensions, acoustic waves incident from any direction will be
completely dissipated.  Below we report two such implementations based
on using DMR's.  In the flat panel composite absorber, a DMR dipolar
resonator was built on the same panel with a pair of DMRs that are
coupled by a thin layer of sealed air, which can generate a monopole
resonance as well as a dipole resonance \cite{yang2013coupled}.  When
the coupled-DMR's monopolar resonance has the same resonance frequency
as the DMR, perfect absorption of sounds is observed in both numerical
simulations and experiment, provided the value of the absorption
coefficient is within a suitable range.  In the second implementation,
the ventilated composite absorber, we place a DMR in the center of a
hollow tube and mount a HMR on the sidewall to generate monopolar
resonance.  When the HMR resonates at the same frequency of the DMR's
Fano resonance
\cite{goffaux2002evidence,fang2006ultrasonic,luk2010fano,ma2013low},
perfect absorption has also been found while the air can freely flow
through the tube.

The basis of understanding the perfect absorption by this type of
degenerate resonators is that, for a resonator in a planar array with
sub-wavelength dimension, only its surface-averaged displacement
$\langle W\rangle$ over the unit cell, i.e., piston-like motion
component, couples to radiative modes in air.  Here, $W$ denotes the
resonator's displacement normal to the plane and the brackets denote
surface averaging. The remaining component of the displacement,
$\delta W=W-\langle W\rangle$, generates only evanescent waves and
hence can be characterized as `deaf'.  The process of scattering sound
by a DMR is thereby one-dimensional
\cite{yang2008membrane,yang2013coupled,ma2014acoustic}, and for time
harmonic waves the impedance $Z$ is inversely related to the Green
function $G$, i.e., $Z=i/(\omega G)$, where $G=\langle
W\rangle/\langle p\rangle$, with $p$ denoting pressure.

For a monopole resonator and a dipole resonator placed side by side on
a flat surface, the surface-averaged displacement $\langle W\rangle$
on the transmission side, for the two resonators combined, would
vanish if $G_d-G_m=0$, i.e., the two responses cancel each other
through destructive interference (refer to Appendix
\ref{sec:app_absorption} Here the subscripts $d$ and $m$ denote dipole
and monopole, respectively.  As the monopole resonator has two
membranes oscillating out of phase, whereas the dipole resonator has
only one membrane, it must be the case that the dipole resonator is
out of phase of one of the monopole resonator's surface.  Hence the
above condition should always be possible to be satisfied.  For the
incident wave side, the two resonators' responses are in phase.  Here
the backscattering is eliminated through impedance matching to air.
As $Z\equiv p/\langle\dot W\rangle=i/(\omega G)$, we must have
$(G_m+G_d)/2=i/(\omega Z_0)$, where $Z_0$ denotes the air impedance.
Therefore, to completely eliminate scatterings, we should have,
\begin{equation}
	G_m=i/(\omega Z_0)=G_d.
	\label{eq:ab_condition}
\end{equation}
While the above description uses the flat composite absorber as the
concrete example, exactly the same applies to the ventilated composite
absorber.

The response functions in Eq.~\eqref{eq:ab_condition} can be written
explicitly in terms of eigenmodes
\cite{yang2013coupled,yang2014homogenization}.  In the vicinity of a
monopolar resonance, the associated $G_m$ is given by the Lorentzian
form,
\begin{equation}
	G_{m}=\frac
	{|\langle W_{m}^f\rangle-\langle W_{m}^b\rangle|^2}
	{2\rho_{m}[(\omega_{m}^2-\omega^2)^2+\omega^2\beta_{m}^2]}
	[(\omega_m^2-\omega^2)+i\omega\beta_m],
	\label{eq:green_m}
\end{equation}
where the superscripts $f$ and $b$ on $W$ indicate the front and back
surfaces, $\rho_{m}\equiv\int_\Omega\rho|W_{m}|^2dV$ is a parameter
related to the displacement-weighted mass density for the monopolar
eigenmode $W_{m}$ resonating at $\omega_{m}$, $\rho$ is the local mass
density, and $\Omega$ the volume of the resonator.  If we denote the
viscosity coefficient of the system as $\eta$
\cite{landau1970elasticity}, the dissipation coefficient $\beta_{m}$
in Eq.~\eqref{eq:green_m} is defined by \cite{yang2015sound}:
\begin{equation}
	\beta_m=\int_\Omega\eta|\nabla(\delta W_m)|^2
	dV/\rho_m,
	\label{eq:dissipation}
\end{equation}
where $\eta$ can be treated as a fitting parameter from experimental
testing.  Similarly, the dipolar response $G_d$ is given by
\begin{equation}
	G_{d}=\frac
	{2|\langle W_{d}\rangle|^2}
	{\rho_{d}[(\omega_{d}^2-\omega^2)^2+\omega^2\beta_{d}^2]}
	[(\omega_d^2-\omega^2)+i\omega\beta_d].
	\label{eq:green_d}
\end{equation}

Note that Eq.~\eqref{eq:ab_condition} requires both responses to be
imaginary.  According to Eqs.~\eqref{eq:green_m} and
\eqref{eq:green_d}, this can be fulfilled when the monopole and dipole
resonances are degenerate at the same frequency, so that at
$\omega=\omega_m=\omega_d$, $\text{Re}(G_m)=\text{Re}(G_d)=0$.  By
adjusting their dissipation coefficients $\beta_{m(d)}$, such as
through the intensity of the `deaf' components $\delta W_{m(d)}$,
perfect absorption can be achieved when the two modes cancel each
other at the transmission side and match the impedance of air at the
backscattering direction.

Normalized to incoming wave, energy absorbed by such a composite wave
absorber is given by (with details shown in Appendix
\ref{sec:app_absorption}).
\begin{align}
	\nonumber
	A=&\frac{2\omega Z_0\text{Im}(G_m)}
	      {[1+\omega Z_0\text{Im}(G_m)]^2+\omega^2 Z_0^2 \text{Re}(G_m)^2}\\
	&+\frac{2\omega Z_0\text{Im}(G_d)}
	      {[1+\omega Z_0\text{Im}(G_d)]^2+\omega^2 Z_0^2 \text{Re}(G_d)^2},
	\label{eq:absorption}
\end{align}
which comprises two terms from the monopolar and dipolar resonances,
respectively.  Each term reaches a maximum of 50\% when
Eq.~\eqref{eq:ab_condition} is satisfied, similar to the CPA
conditions \cite{wei2014symmetrical,yang2015sound}.

\begin{figure}
\includegraphics[scale=0.90]{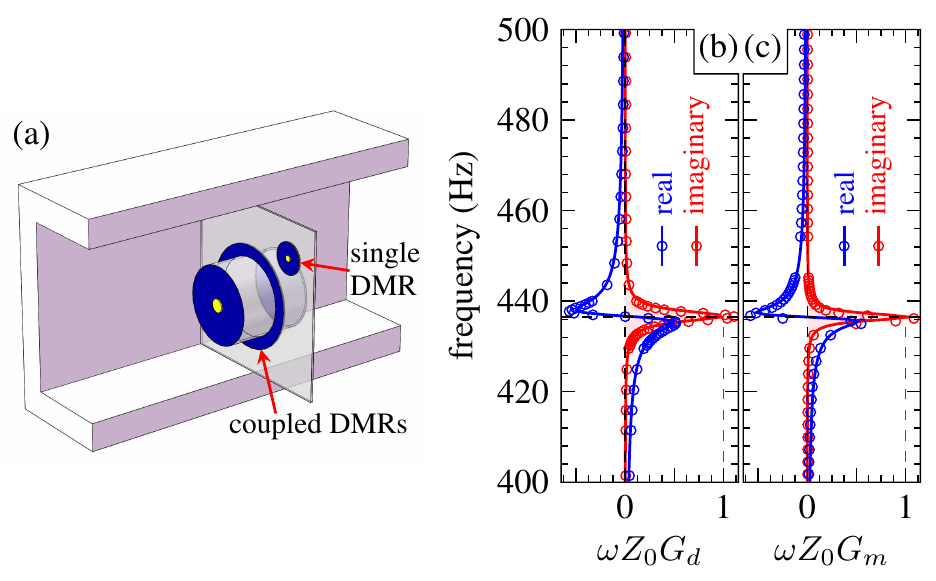}
\caption{
	(a) Schematic cutoff view of the flat panel composite absorber.
	Based on the parameters in Table~\ref{tab:parameter}, the
	associated responses are given by Eqs.~\eqref{eq:green_m} and
	\eqref{eq:green_d} as shown in (b) for the dipolar response and
	(c) for the monopole response.  Solid curves stand for theory, and
	open circles are predictions based on parameters retrieved from
	experiments.
}
\label{fig:structure}
\end{figure}

Our idea of realizing such a degenerate perfect composite absorber is
to assemble a monopolar and a dipolar resonator together while keeping
the overall size in the subwavelength regime
\cite{ding2007metamaterial,lee2010composite}. The absorption unit in
the flat panel composite absorber is a 90$\times$90 mm panel
consisting of a single DMR for dipolar resonance and a pair of coupled
DMRs for monopolar resonance as shown in Fig.~\ref{fig:structure}(a).
Stretched on a rigid circular frame with radius of 8 mm, the single
DMR is a 0.2 thick rubber membrane (blue circle) attached by a 56 mg
weight (yellow button) in center.  In the coupled-DMR, two 40-mm wide
identical DMRs [blurred circles in Fig.~\ref{fig:structure}(a)] seal a
rigid cylindrical tube (gray cylinder) on the two ends.  In the
monopole resonance mode, the two DMRs, each with a 70 mg button
attached, oscillate out of phase with each other and thereby
compressing and expanding the air layer in-between.  The coupled-DMR
resonator is isolated from the nearby dipole resonator, situated
$\sim$1 cm away, by a ring-shaped membrane [blue ring in
Fig.~\ref{fig:structure}(a)].  Numerical simulations by COMSOL
Multiphysics (with material parameters given in Ref.
[\onlinecite{yang2014homogenization}]) indicate a pair of almost
degenerate resonances for DMR's dipole resonance at 436.7 Hz and the
coupled-DMR's monopolar resonance at 436.5 Hz.

\begin{table}
\begin{tabular}{ccccc}
\hline\hline
										&\#1: $G_m$	&\#1: $G_d$	&\#2: $G_m$	&\#2: $G_d$	\\
\hline
$\alpha$ (m$^2$/kg) &0.031			&0.043			&0.11				&4.93				\\
$\beta$ (Hz)				&12.3				&17.8				&6.8				&8.0				\\
$\eta$ (Pa$\cdot$s) &1.14				&1.18				&0.95				&0.722			\\
\hline\hline
\end{tabular}
\caption{
	Evaluated parameters for each response from numerical simulations. 
	$\alpha=|\langle W_m^f\rangle-\langle W_m^b\rangle|^2/(2\rho_m)$
	for $G_m$ in the flat panel composite absorber (\#1), $\alpha=|\langle
	U_m\rangle|^2/\rho_m$ for $G_m$ in the ventilated composite
	absorber (\#2), and	$\alpha=2|\langle W_d\rangle|^2/\rho_d$ for
	$G_d$.
}
\label{tab:parameter}
\end{table}

We plot the responses of the two resonators, in the form of their
Green functions (normalized by $\omega Z_0$) as functions of frequency
(solid colored curves) in Figs.~\ref{fig:structure}(b,c), in which
parameters in Eqs.~\eqref{eq:green_m} and \eqref{eq:green_d} are
evaluated in Table~\ref{tab:parameter} based on simulations.  By
experimentally measuring the reflection $R$ and transmission $T$ for
each resonator, their response function can also be retrieved through
$\omega Z_0G_{m(d)}=i[1-(R\pm T)]/[1+(R\pm T)]$ (Appendix
\ref{sec:app_retrieval}), as shown by open circles in
Figs.~\ref{fig:structure}(b,c).  It should be especially noted that at
the resonances, Eq.~\eqref{eq:ab_condition} has been fulfilled with
$\omega Z_0G_m=1.133i$ and $\omega Z_0G_d=1.105i$, i.e., impedance
matched.

\begin{figure}
\includegraphics[scale=0.9]{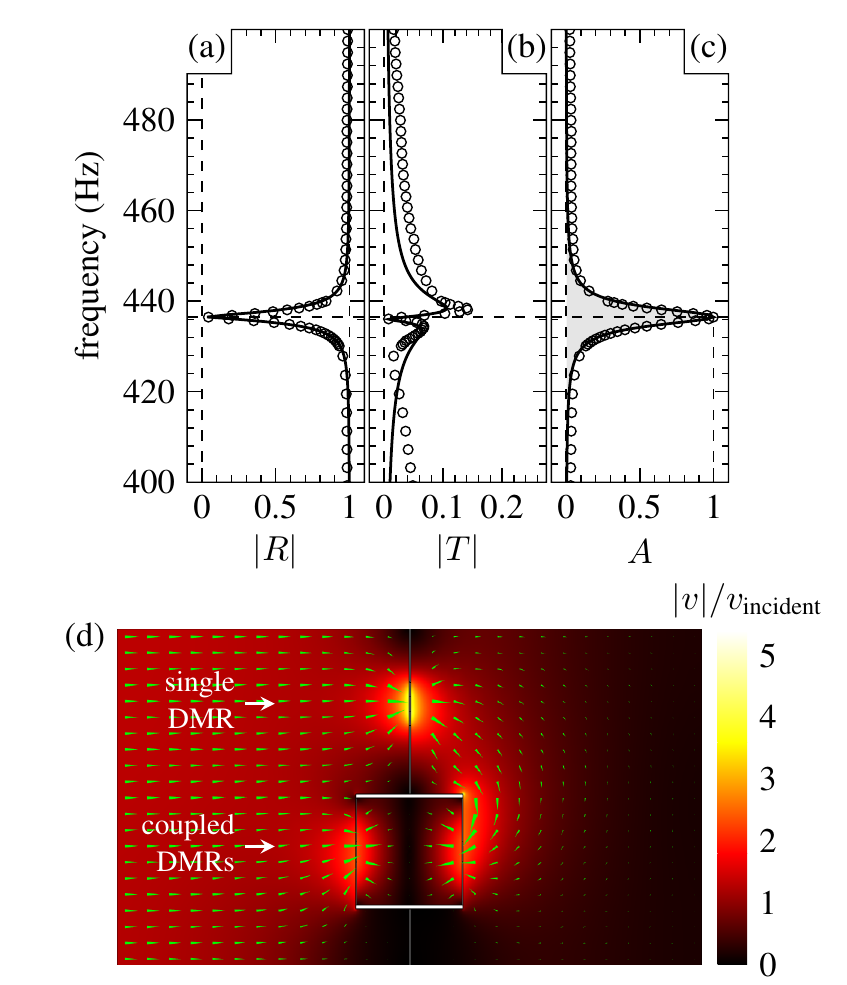}
\caption{
	(a-c) Scatterings and absorption coefficient of the flat panel
	composite absorber.  Almost perfect absorption is seen at 436.5
	Hz.  The solid curves are from theory and the open circles from
	experiments.  (d) Cross-sectional profile of air velocities at
	436.5 Hz from numerical simulations.  The small arrows indicate
	the in-plane velocity with length proportional to its magnitude.
	The acoustic wave is incident from the left.
}
\label{fig:absorption}
\end{figure}

In the experiment, the sample is sandwiched between two impedance
tubes. With plane wave generated by the speaker at one end, scattered
waves from the sample were measured by sensors mounted in the tubes on
both sides of the sample.  Details of measurements are shown in
Appendix~\ref{sec:experiment}.  The results for this composite
absorber are shown in Fig.~\ref{fig:absorption}.  As expected, an
absorption peak reaching 99.7\% is seen at the degenerate resonance
frequency of 436.5 Hz [Fig.~\ref{fig:absorption}(c)].  The associated
reflection and transmission coefficients are shown in
Figs.~\ref{fig:absorption}(a,b).  Very good agreement between
experiment (open circles) and theory (solid curves) is seen.  The
relevant airborne wavelength, 786 mm, is noted to be more than 20
times that of the absorber's thickness, and 10 times larger than its
width.  The simulated profile of air velocity is shown in
Fig.~\ref{fig:absorption}(d).  On the transmission side the air
velocities in the vicinity of the single DMR are seen to have the
opposite symmetry from that of the coupled-DMR, thereby canceling each
other further away.  On the incident side the plane wave is seen to
maintain its amplitude, implying no backscattered wave owing to
matched impedance.

\begin{figure}
\includegraphics[scale=0.9]{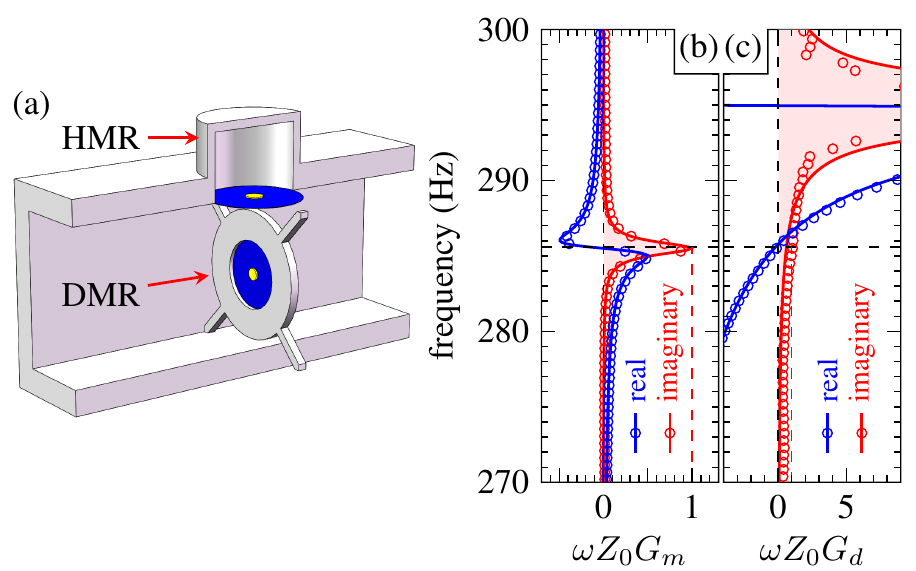}
\caption{
	(a) A schematic cutoff view of the ventilated composite absorber
	that comprises a sidewall monopole resonator placed in the
	vicinity of dipole resonator.  The relevant responses are given by
	Eqs.~\eqref{eq:green_m2} and \eqref{eq:green_fano}.  They are shown
	by solid curves in (b) for monopole sidewall resonator and in (c)
	for the dipolar resonator.  Predictions with parameter values
	retrieved from experiment are shown by open circles.  Very good
	agreement is seen.  Note the Fano profile of $G_d$ in (c)
}
\label{fig:structure2}
\end{figure}

In the ventilated composite absorber, shown in
Fig.~\ref{fig:structure2}(a), a sub-wavelength short tube having a
90$\times$90 mm square cross-section is seen to have a hybrid membrane
resonator (HMR) attached on the sidewall.  The HMR is a 35 mm deep, 55
mm wide, cylindrical chamber sealed by a rubber membrane (blue
circle), with a 130 mg attached weight (yellow button).  Like the
sidewall Helmholtz resonator \cite{fang2006ultrasonic}, the HMR
generates a monopole response in air, along the tube's axial
direction, through the sucking and pushing of air associated with its
membrane's normal displacement.  The effect on the two ports (two
sides of the DMR) is proportional to $f_m/2$, where $f_m=0.29$ is the
areal ratio between the membrane and the tube's cross-sections.
Similar to Eq.~\eqref{eq:green_m} \cite{ma2014acoustic}, $G_m$ is
given by
\begin{equation}
	G_m=\frac{
	f_m|\langle U_{m}\rangle|^2/2}
	{\rho_{m}[(\omega_{m}^2-\omega^2)^2+\omega^2\beta_{m}^2]}
	[(\omega_m^2-\omega^2)+i\omega\beta_m],
	\label{eq:green_m2}
\end{equation}
where $U_m$ is the HMR's  membrane normal displacement field at the
hybrid resonance with $\omega_m=2\pi\times285.3$ Hz.  Based on the
values in Table~\ref{tab:parameter}, Eq.~\eqref{eq:green_m2} gives the
solid curves in Fig.~\ref{fig:structure2}(b).  Experimental results
(open circles) are seen to agree with the theory very well.  Moreover,
at the resonance frequency the data are seen to agree with the
prediction of Eq.~\eqref{eq:ab_condition}, with $\omega Z_0G_m=0.93i$.

The dipole resonator, a 40 mm-wide DMR (blue circle), is placed at the
center of the square tube, with a rigid rim (gray ring) that is 8 mm
in its width.  The whole structure is supported by four poles
[Fig.~\ref{fig:structure2}(a)].  We note that the membrane and its rim
block the tube's cross-section with a ratio $f_d=0.55$.  While the DMR
gives rise to narrow discrete resonances, the air in the remaining
cross-section area contributes a smooth continuum background, with an
extra term $-(1-f_d)/(\tau\rho_0\omega^2)$ in $G_d$ \cite{ma2013low},
so that
\begin{align}
	\nonumber
	G_d=&f_d\frac
	{2|\langle W_{d}\rangle|^2}
	{2\rho_{d}[(\omega_{d}^2-\omega^2)^2+\omega^2\beta_{d}^2]}
	[(\omega_d^2-\omega^2)+i\omega\beta_d]\\
	&-(1-f_d)\frac{1}{\tau\rho_0\omega^2}.
	\label{eq:green_fano}
\end{align}
Here $\tau$(=8 mm) denotes width of the ring,  In contrast to the
Lorentzian resonance in Eq.~\eqref{eq:green_d}, this response exhibits
a distinctly asymmetric-shaped Fano resonance due to the interference
between the two terms in Eq.~\eqref{eq:green_fano}, at which another
zero $\text{Re}(G_d)$ exists below $\omega_d$
\cite{goffaux2002evidence,fang2006ultrasonic,luk2010fano,ma2013low}.
To make it degenerate with the HMR, we adjust the weight on the DMR
(yellow button) to be 38 mg (so that $\omega_d=2\pi\times295.0$ Hz).
Based on the parameters in Table~\ref{tab:parameter},
Eq.~\eqref{eq:green_fano} yields a dipolar response (solid curves),
shown in Fig.~\ref{fig:structure2}(c), that agrees well with the
prediction obtained with experimentally retrieved parameter values
(open circles).  At the degenerate resonance frequency we have $\omega
Z_0G_d=1.07i$, again agreeing with the condition given by
Eq.~\eqref{eq:ab_condition} reasonably well.

\begin{figure}
\includegraphics[scale=0.9]{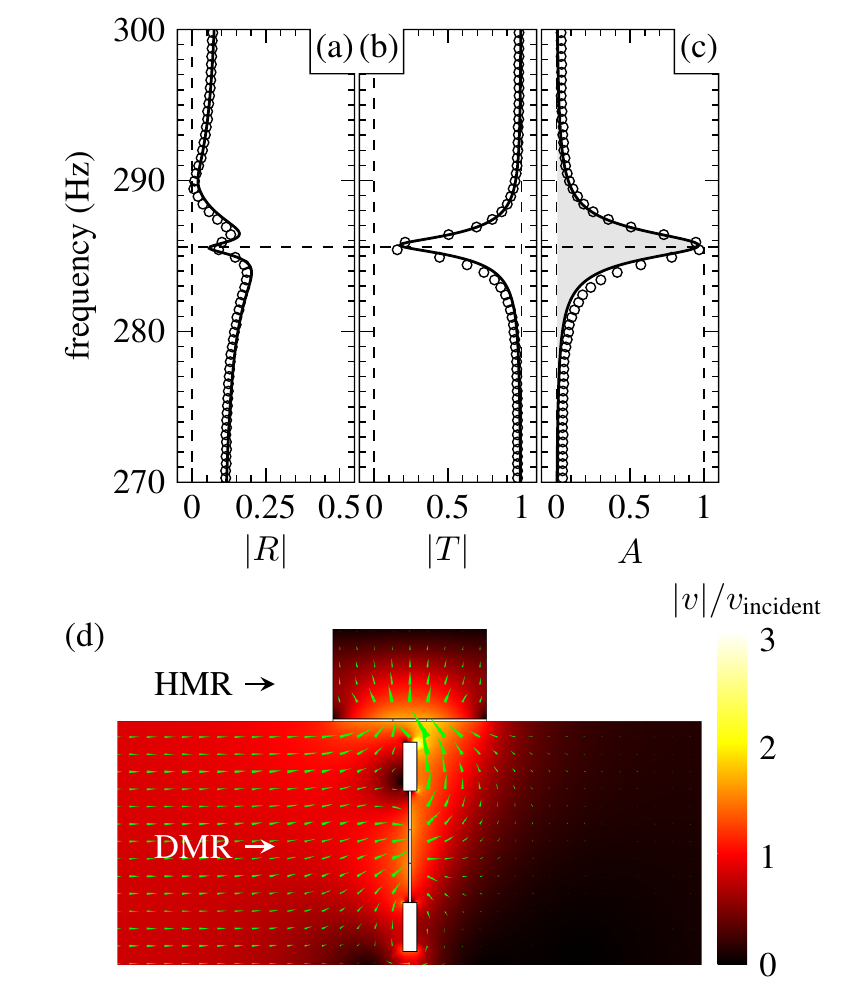}
\caption{
	(a-c) The reflection, transmission, and absorption coefficients
	for the ventilated composite absorber.  Almost perfect absorption,
	reaching 99.2\%, is seen at 285.6 Hz.  The solid curves are from
	theory and the open circles from the experiments.  (d) Simulated air
	velocities at the total absorption frequency.  The green arrows
	are in-plane velocity with its length being proportional to the
	magnitude.  The acoustic wave is incident from the left.
}
\label{fig:absorption2}
\end{figure}

The combined effect of monopole and dipole resonances is shown in
Figs.~\ref{fig:absorption2}(a-c).  The composite absorber exhibits
almost perfect absorption of 99.2\% at 285.6 Hz, with vanishing
reflection and transmission.  The relevant airborne wavelength is
about 1.2 m, i.e., 22 times larger than the absorber's thickness and
10 times larger that its width.  Figure~\ref{fig:absorption2}(d) shows
the simulated air velocities at the peak absorption frequency.  Again,
the air motion is cancelled on the transmission side, with impedance
matching behavior on the incident side.  It should be noted that the
near-total absorption is achieved in this case while the airflow can
still flow largely unimpeded in the channel.

In summary, we have shown that by combining a pair of degenerate
monopole and dipole resonators with subwavelength dimensions, perfect
absorption of sound can be achieved.  The subwavelength dimension of
the absorber unit implies that the absorption functionality is
independent of the incident direction.  We have experimentally
demonstrated the effect by two examples: the flat panel absorber with
single DMR mounted in the same panel as a coupled-DMR, and the
ventilated absorber comprising a short tube with HMR mounted on its
sidewall and a DMR placed in the center.  Almost perfect absorption
for sounds with wavelength at least 10 times larger than the absorber
has been observed in both cases, with excellent agreement between
theory and experiment.  In addition, we would like to note that, owing
to the similarity between acoustic and electromagnetic waves, the
present proposed total absorption mechanism by degenerate
mirror-symmetric and anti-symmetric resonances should be valid for
electromagnetic waves as well, with specific polarizations.

M.Y. wishes to thank Guancong Ma for helpful discussions.  This work is
supported by AoE/P-02/12 from the Research Grant Council of the Hong
Kong SAR government.

\appendix
\section{Scattering Amplitudes and Absorption Coefficient}
\label{sec:app_absorption}

Consider a monopole resonator at $z=0$.  It scatters plane acoustic
wave incident from two sides, with two outgoing waves.  With $k_0$
being acoustic wavevector in air, we denote the incoming waves as
$p=\exp(ik_0z)$ when $z<0$ and $p=\exp(-ik_0z)$ when $z>0$, and the
symmetric outgoing waves as $p=s_m\exp(-ik_0z)$ for $z<0$ and
$p=s_m\exp(ik_0z)$ for $z>0$.  Here $s_m$ denotes the scattering
coefficient.  The resonator's monopole response on two sides can be
characterized by the symmetric displacement $W$ that is proportional
to the applied pressure via the Green function $G_m$, i.e.,
$W^b=-W^f=G_m(1+s_m)$, here the thickness of $2\epsilon$ for the
resonator is assumed to be small compared to the wavelength so that
$k_0\epsilon\ll1$.  Superscripts $b$ and $f$ indicate back and front
surfaces, respectively.  From Newton's law, the amplitude of air
compression (and expansion) between the two surfaces is related to
pressure by the relationship $W=(\partial p/\partial
z)/(\omega^2/\rho_0)$.  It follows that $W^b=-W^f=\xi(1-s_m)$.  Here
$\rho_0$ is the density of air and $\xi\equiv i/(\omega Z_0)$, with
$Z_0$ being the air impedance.  From the continuity of displacement,
the scattering coefficient $s_m$ is given by
\begin{equation}
  s_m=\frac{\xi-G_m}{\xi+G_m}.
  \label{eq:monopole_s}
\end{equation}

Similarly, a dipole resonator scatters the incident waves
$p=\exp(ik_0z)$ for $z<0$ and $p=-\exp(-ik_0z)$ for $z>0$ into
anti-symmetric outgoing waves with $p=s_d\exp(-ik_0z)$ for $z<0$ and
$p=-s_d\exp(ik_0z)$ for $z>0$, where $s_d$ is given by
\begin{equation}
  s_d=\frac{\xi-G_d}{\xi+G_d}.
  \label{eq:dipole_s}
\end{equation}

For a composite resonator comprising monopole and dipole resonances
with wave incident from only one side, so that $p=\exp(ik_0z)$ when
$z<0$ and $p=0$ when $z>0$, the reflection $R$ can be obtained by
simply superposing the two aforementioned cases:
\begin{equation}
  R=\frac{1}{2}(s_m+s_d)=\frac{\xi^2-G_dG_m}{(G_d+\xi)(G_m+\xi)}.
  \label{eq:r}
\end{equation}
The transmission can be similarly obtained as
\begin{equation}
  T=\frac{1}{2}(s_m-s_d)=\frac{(G_d-G_m)\xi}{(G_d+\xi)(G_m+\xi)}.
  \label{eq:t}
\end{equation}
Since $A=1-|R|^2-|T|^2$, the absorption coefficient $A$ is therefore
given by
\begin{align}
	\nonumber
	A=&\frac{2\omega Z_0\text{Im}(G_m)}
	      {[1+\omega Z_0\text{Im}(G_m)]^2+\omega^2 Z_0^2 \text{Re}(G_m)^2}\\
	&+\frac{2\omega Z_0\text{Im}(G_d)}
	      {[1+\omega Z_0\text{Im}(G_d)]^2+\omega^2 Z_0^2 \text{Re}(G_d)^2},
	\label{eq:a}
\end{align}

\section{Retrieval of Responses from Scatterings}
\label{sec:app_retrieval}

It is straightforward to retrieve the monopole and dipole responses,
$G_m$ and $G_d$, from the scattering coefficients $R$ and $T$.  As
$s_m=R+T$ and $s_d=R-T$, the two responses can be directly solved from
Eqs.~\eqref{eq:monopole_s} and \eqref{eq:dipole_s}:
\begin{align}
	G_m=\frac{1-(R+T)}{1+(R+T)}\xi,\quad
	G_d=\frac{1-(R-T)}{1+(R-T)}\xi.
	\label{eq:retrieval}
\end{align}
As $R$ and $T$ can be measured experimentally, substituting their
values into Eq.~\eqref{eq:retrieval} directly yields the relevant
monopole and dipole responses.

\section{Experiment Setup and the Measurement Method}
\label{sec:experiment}

\begin{figure}
\centering
\includegraphics[]{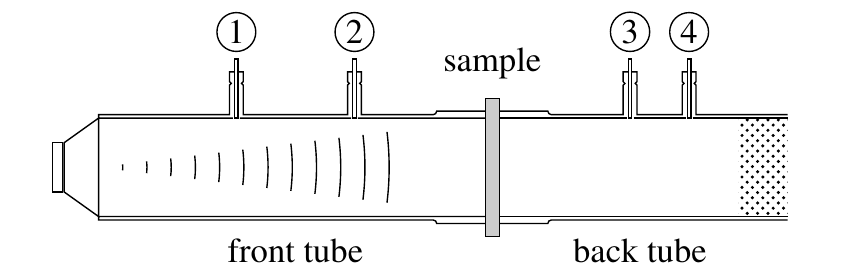}
\caption{
  Schematic illustration of the experimental apparatus.  The front
  tube with a loudspeaker at one end has two sensors, labeled by 1 and
  2, while the back tube, plugged by acoustic foam at the end, has two
  more sensors, as labeled by 3 and 4.
}
\label{fig:setup}
\end{figure}

We test the sample's scattering and absorption properties by
sandwiching it between two impedance tubes having square
cross-sections.  The front tube has two sensors
[Fig.~\ref{fig:setup}], plus a loudspeaker at the front end to
generate the plane waves. The back tube has another two sensors, and
the tube's back end is filled by acoustic foam to eliminate
reflection.  By normalizing the pressure amplitude of all the relevant
sound waves by the incident sound pressure amplitude, the reflection
and transmission coefficients, $R$ and $T$, can be obtained from the
pressure data recorded by the four sensors \cite{yang2015sound}.

As scalar waves, airborne sound can propagate in a sub-wavelength
waveguide without a cut-off frequency. In our experiments, the
geometrical size of the apparatus (viz., the width of the square
waveguide) is smaller than the
relevant wavelength, so that only plane waves can propagate in both
the front and back tubes \cite{morse01}.  The total pressure fields in
the two (front and back) impedance tubes may be expressed as the sum
of forward and backward waves propagating along the $z$ direction:
\begin{subequations}
\label{eq:Total_pressure}
	\begin{align}
	&p_1=p_1^\text{i}{e^{i{k_0}z}} + p_1^\text{o}{e^{ - i{k_0}z}},\\
	&p_2=p_2^\text{i}{e^{-i{k_0}z}} + p_2^\text{o}{e^{i{k_0}z}}.
	\end{align}
\end{subequations}
Here the subscripts ``1'' and ``2'' refer to the front and back tubes,
and the superscripts ``i'' and ``o'' denote the incoming and outgoing
waves, respectively.  To retrieve the transmission and reflection
coefficients, the total pressure field should first be exactly expressed,
by using the four experimentally measured parameters: $p_1^\text{i},
p_1^\text{o}, p_2^\text{i}$ and $p_2^\text{o}$.  Four sensors are used
to determine these parameters.  Two sensors labeled ``1" and ``2" are
placed in the front tube at $z_1=-339.5$ mm and $z_2=-239.5$ mm, and
the other two sensors labeled ``3" and ``4" are placed in the back
tube at $z_3=193.0$ mm and $z_4=393.0$ mm [cf. Fig.~\ref{fig:setup}].
Here $z=0$ is the sample position. From Eq.~\eqref{eq:Total_pressure},
the pressure values at the positions of the four sensors are 
\begin{subequations}
\label{eq:pressure_sensors}
	\begin{align}
	&p(z_1)=p_1^\text{i}{e^{i{k_0}z_1}} + p_1^\text{o}{e^{-i{k_0}z_1}},\\
	&p(z_2)=p_1^\text{i}{e^{i{k_0}z_2}} + p_1^\text{o}{e^{-i{k_0}z_2}},\\
	&p(z_3)=p_2^\text{i}{e^{-i{k_0}z_3}} + p_2^\text{o}{e^{i{k_0}z_3}},\\
	&p(z_4)=p_2^\text{i}{e^{-i{k_0}z_4}} + p_2^\text{o}{e^{i{k_0}z_4}}.
	\end{align}
\end{subequations}
By solving Eq.~\eqref{eq:pressure_sensors}, we obtain
\begin{subequations}
\label{eq:parameters}
	\begin{align}
	&p_1^\text{i}=\frac{p(z_1) e^{i{k_0}z_1}-p(z_2) e^{i{k_0}z_2}}{e^{2ik_0 z_1}-e^{2ik_0 z_2}},\\
	&p_1^\text{o}=-\frac{p(z_1) e^{i{k_0}z_2}-p(z_2) e^{i{k_0}z_1}}{e^{2ik_0 z_1}-e^{2ik_0 z_2}}e^{ik_0(z_1+z_2)} ,\\
	&p_2^\text{i}=-\frac{p(z_3) e^{i{k_0}z_4}-p(z_4) e^{i{k_0}z_3}}{e^{2ik_0 z_3}-e^{2ik_0 z_4}}e^{ik_0(z_3+z_4)} ,\\
	&p_2^\text{o}=\frac{p(z_3) e^{i{k_0}z_3}-p(z_4) e^{i{k_0}z_4}}{e^{2ik_0 z_3}-e^{2ik_0 z_4}},	
	\end{align}
\end{subequations}
Here $p(z_j)$ is the pressure measured by each sensor
labeled as ``$j=1\sim4$" in the subscripts.

The scattering matrix $S(k_0)$ describing the relationship between the
incoming and outgoing waves can be expressed as 
\begin{equation}
	\label{eq:Smatrix}
	\left( {\begin{array}{*{20}{c}}
		{p_2^\text{o}}\\
		{p_1^\text{o}}
	\end{array}} \right) = S(k_0)\left( 
	{\begin{array}{*{20}{c}}
		{p_1^\text{i}}\\
		{p_2^\text{i}}
	\end{array}} \right),
	\quad S(k_0) = \left( 
	{\begin{array}{*{20}{c}}
		T&R\\
		R&T
	\end{array}} \right).
\end{equation}
It should be noted that, due to the symmetry of the sample in our
system, the reflection and transmission coefficients $R$ and $T$ are
identical if the sample is tuned around 180 degrees.  As a result, the
reflection and transmission coefficients can be retrieved as
\begin{subequations}
\label{eq:RTcoeff}
\begin{align}
	&R=\frac{p_1^\text{i} p_1^\text{o} - p_2^\text{i}
	p_2^\text{o}}{p_1^\text{i} p_1^\text{i} - p_2^\text{i}
	p_2^\text{i}},\\
	&T=\frac{p_1^\text{i} p_2^\text{o} - p_1^\text{o}
	p_2^\text{i}}{p_1^\text{i} p_1^\text{i} - p_2^\text{i} p_2^\text{i}}.
\end{align}
\end{subequations}
Here the four wave amplitude $p_{1}^\text{i}$, $p_{1}^\text{o}$,
$p_{2}^\text{i}$ and $p_{2}^\text{o}$ are determined from
Eq.~\eqref{eq:parameters}.

\bibliography{main}

\end{document}